# Surface structures of ZrO$_2$ films on Rh(111): From two layers to bulk termination

Peter Lackner[1], Zhiyu Zou[1], Sabrina Mayr[1], Joong-Il Jake Choi[1,^], Ulrike Diebold[1], and Michael Schmid[1]*

[1] Institute of Applied Physics, TU Wien, Vienna, Austria

[^] Present address: Center for Nanomaterials and Chemical Reactions, Institute for Basic Science (IBS), Daejeon 305-701, South Korea.

*email address: schmid@iap.tuwien.ac.at

## Abstract

We have studied zirconia films on a Rh(111) substrate with thicknesses in the range of 2–10 monolayers (ML) using scanning tunneling microscopy (STM) and low-energy electron diffraction (LEED). Zirconia was deposited using a UHV-compatible sputter source, resulting in layer-by-layer growth and good uniformity of the films. For thicknesses of 2–4 ML, a layer-dependent influence of the substrate on the structure of the thin films is observed. Beyond this thickness, films show a (2 × 1) or a distorted (2 × 2) surface structure with respect to cubic ZrO$_2$(111); these structures correspond to tetragonal and monoclinic zirconia, respectively. The tetragonal phase occurs for annealing temperatures of up to 730 °C; transformation to the thermodynamically stable monoclinic phase occurs after annealing at 850 °C or above. High-temperature annealing also breaks up the films and exposes the Rh(111) substrate. We argue that the tetragonal films are stabilized by oxygen deficiency, while the monoclinic films are only weakly defective and show band bending at defects and grain boundaries. This observation is in agreement with positive charge being responsible for the grain-boundary blocking effect in zirconia-based solid electrolytes. Our work introduces the tetragonal and monoclinic 5 ML-thick ZrO$_2$ films on Rh(111) as well-suited model system for surface-science studies on ZrO$_2$ as they do not exhibit the charging problems of thicker films or the bulk material and show better homogeneity and stability than the previously-studied ZrO$_2$/Pt(111) system.

# 1. Introduction

The search for a detailed understanding of a material is often driven by the technological applications that rely on it. This is also true for zirconia ($ZrO_2$), which is used as catalyst support [1] and catalyst [2], as a refractory ceramic due to its high thermal stability and strength [3–5], and as dental implant material due to its good biocompatibility [6]. Chemically doped zirconia is heavily used as a solid-state electrolyte in solid oxide fuel cells [7] and gas sensors [8]. While the material is an electronic insulator up to high temperatures, it can conduct oxygen (and, thereby, electric charge) via vacancy diffusion, which forms the basis for using zirconia as an electrolyte. As the intrinsic concentration of oxygen vacancies ($V_O$s) in $ZrO_2$ is very low even at high temperatures and reducing atmosphere, $V_O$s are introduced by chemical doping with trivalent elements such as yttrium.

Depending on temperature or dopant concentration, zirconia exhibits three stable bulk structures at atmospheric pressure: For pure, stoichiometric $ZrO_2$, the cubic structure (c-$ZrO_2$, fluorite lattice) is found above 2377 °C; at lower temperatures the tetragonal phase (t-$ZrO_2$, above 1205 °C), and finally monoclinic $ZrO_2$ (m-$ZrO_2$, also known as baddeleyite) are stable [9], see Figure 1a. While all these phases are related to the cubic fluorite structure, pure c-$ZrO_2$ does not exist at room temperature due to the small O–O distance imposed by the short (strong) Zr–O bonds ($d_{O-O} \approx 256$ pm for hypothetical room-temperature c-$ZrO_2$ [10]). The average O–O distance can be increased by shifting the O atoms alternatingly up or down in [001] direction, leading to the tetragonal phase ($d_{O-O} \approx 260$ pm), which, however, is still unstable for pure, stoichiometric $ZrO_2$ at room temperature. Upon transformation to m-$ZrO_2$, Zr–O bonds are broken, the coordination of Zr changes from 8 to 7 and for half of the O atoms the coordination is reduced from fourfold (tetrahedral) to threefold (planar); the volume increases by $\approx 5\%$. These changes substantially increase the average O–O distance, while the Zr–O bonds remain short.

The tetragonal and cubic phases can be also stabilized at room temperature by introducing $V_O$s and by increasing the lattice parameter, which occurs when doping with yttria [9,11–13]. $Y_2O_3$ concentrations above $\approx 8$ mol% (corresponding to $Zr_{0.85}Y_{0.15}O_{1.93}$) stabilize the cubic phase (yttria-stabilized zirconia, YSZ). At lower dopant concentrations one finds mixtures of cubic and tetragonal, or cubic and monoclinic zirconia. The lower doping limit is 1.5–2 mol% $Y_2O_3$, where the monoclinic phase becomes stable [14]. The tetragonal phase is also found in nanoscale $ZrO_2$ at room temperature; while this was initially attributed to its favorable surface energy [15], newer works rather point towards a stabilization by $V_O$s instead [12,13].

Figure 1b shows the surface termination of c-ZrO$_2$(111) and the corresponding lowest-energy terminations of the other phases. Despite the distortions with respect to c-ZrO$_2$, all these surfaces are non-polar. The cubic phase exhibits a hexagonal (1 × 1) structure. The shifted O rows in the tetragonal phase lead to a (2 × 1) unit cell w.r.t. the cubic phase. The monoclinic phase features a distorted (2 × 2) surface unit cell (again, w.r.t. c-ZrO$_2$). For this latter phase, density functional theory (DFT) calculations [16] predict that the $(\bar{1}11)$ surface has the lowest surface energy; in contrast to (111), it has only one (instead of two) surface O with twofold coordination per unit cell (marked by an asterisk in Figure 1b).

Apart from the above-mentioned structures, several orthorhombic high-pressure phases of zirconia exist; some of these are metastable at ambient conditions [17]. Recently, the orthorhombic phases of ZrO$_2$ and mixed ZrO$_2$/HfO$_2$ have received increased attention as candidate materials for ferroelectric memory devices [18]. Similar to monoclinic ZrO$_2$, these orthorhombic phases are based on distortions of the cubic fluorite structure, again having 7-fold coordinated Zr and O with 3-fold and 4-fold coordination [19,20]. When cut along a direction equivalent to c-ZrO$_2$(111), the most common orthorhombic I and II phases would exhibit a (2 × 2) or (2 × 4) surface unit cell with respect to the cubic phase, respectively.

In spite of its technological importance, zirconia has received surprisingly little attention from the surface-science community. This is partly due to its insulating nature, as most surface-science methods rely on electronic conductance, e.g., scanning tunneling microscopy (STM), low-energy electron diffraction (LEED), or x-ray photoelectron spectroscopy (XPS). Furthermore, the phase transitions make it impossible to grow a ZrO$_2$ single crystal from the melt. This second limitation does not exist for cubic YSZ, where single crystals are readily available and inexpensive. Morrow *et al.* [21] used a YSZ single crystal for high-temperature STM studies; this work was conducted at ~300 °C to ensure sufficient conductivity. While atomic resolution was achieved, this approach is limited to high temperatures, and due to Y segregation the surfaces had a rather high Y concentration. Several groups have followed a different approach and used several-monolayer-thick films of pure zirconia on Pt(111) as model systems. Meinel *et al.* [22–24] performed STM studies on up to 10 ML-thick zirconia films that were deposited onto Pt(111) via physical vapor deposition (PVD) of Zr in an O$_2$ atmosphere. Their work had been built on a previous LEED study by Maurice *et al.* [25], but involved annealing at higher temperatures. Depending on film thickness and annealing temperature, Meinel *et al.* found a large number of superstructures in LEED and STM. It must be noted that films as thick as 10 ML broke up upon annealing and eventually dissolved in the Pt substrate, so it is not straightforward to decide which structures should be assigned to the multilayer films and which ones to the Pt-Zr or Pt-Zr-O structures. Nevertheless, it is clear that the initial structures were based on ZrO$_2$(111), with a ZrO$_2$(111)-(2 × 2) LEED pattern. Also spots interpreted as $ZrO_2$(111)-(1 × 1) rotated by ±6.6° w.r.t. Pt(111) after 3 min of annealing at 680 °C can be attributed to the $ZrO_2$ films. Submonolayer films exhibited (5 × 5) and ($\sqrt{19} \times \sqrt{19}$)R36.6° superstructures with respect to Pt(111) [26]; the latter structure also appeared when thicker films were annealed at high temperatures. Possibly those formed at areas where the thicker $ZrO_2$ film has disappeared. STM indicates a band gap at least for films ≥ 2 ML and density functional theory (DFT) indicates that bulk-like band gaps are reached at 5 ML [23].

A different approach to zirconia model systems is the growth of ultra-thin zirconia films by oxidation of alloy single crystals as first shown by Antlanger et al. [27]. Two substrates were used: $Pt_3Zr$(0001) [27,28] and $Pd_3Zr$(0001) [29]. By annealing these crystals at 400 °C in $O_2$, disordered zirconia formed, which consumed Zr from the top layers of the alloy. In the case of $Pt_3Zr$, Zr diffusion is slow, so the interface was essentially pure Pt(111). By annealing at 900 °C in UHV, the disordered zirconia transformed to an ordered monolayer corresponding to one O–Zr–O trilayer repeat unit of $ZrO_2$(111). For $ZrO_2/Pt_3Zr$, the film exhibited the same ($\sqrt{19} \times \sqrt{19}$)R23.4° superstructure (this rotation angle is equivalent to 36.6°) as found in previous studies of zirconia on Pt(111) [24,26], confirming that this structure on Pt(111) likely corresponded to an ultrathin film. $ZrO_2/Pd_3Zr$ formed an O–Zr–O trilayer with an almost identical in-plane lattice constant (0.35 nm) and a large ($\sqrt{217} \times \sqrt{217}$)R10.16° superstructure cell. Both, STM measurements and DFT calculations indicated a substantial buckling of the films. These alloy-based, ultrathin zirconia films were successfully used as model systems for surfaces of bulk $ZrO_2$ in metal growth [30] and water adsorption [31] studies.

Thin zirconia films can also be grown by atomic layer deposition using the precursor zirconium (IV) tert-butoxide (ZTB). While this technique is typically used in industry [32] and not in UHV studies, it was successfully applied to deposit sub-monolayer coverages of $ZrO_2$ on Pd(111) [33,34] and Cu(111) [35]. However, it remains to be seen whether this method can be used to grow atomically flat zirconia films with a thickness of a few monolayers.

In the current work, we present results obtained from zirconia films deposited with a UHV-compatible sputter source [36]. This sputter source features a highly reproducible deposition rate, high purity of the films and higher deposition rates than typically achievable by PVD. We found varying structures with increasing film thickness; for thicknesses below 5 ML the films are strongly influenced by the underlying Rh(111) substrate. For thicknesses ≥ 5 ML, bulk-terminated tetragonal or monoclinic zirconia can be stabilized, depending on the annealing temperature.

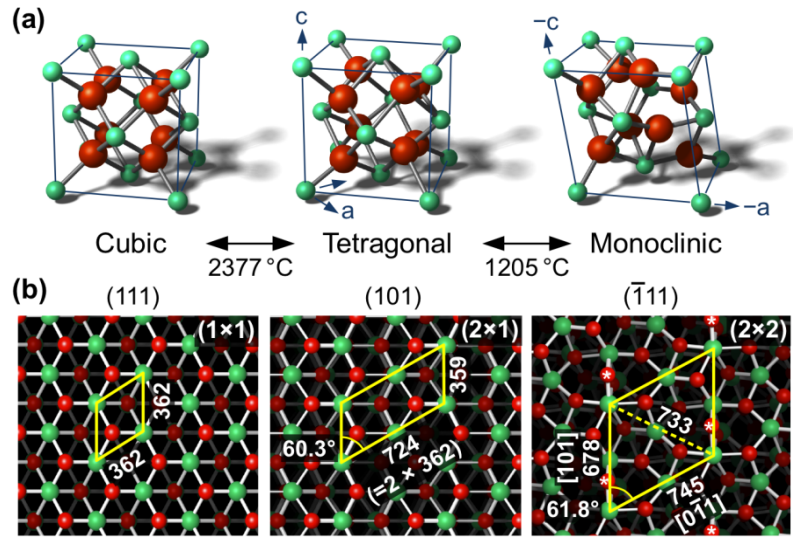

*Figure 1:* The three stable ambient-pressure bulk phases of zirconia. (a) Bulk unit cells and (b) bulk-terminated surfaces equivalent to the (111) surface of cubic $ZrO_2$, with the surface unit cells marked in yellow. Oxygen (2−) ions are depicted in red, zirconium (4+) ions in green. Room-temperature cell sizes in (b) are given in pm and based on Ref. [10,37] for cubic (extrapolated from doped to undoped $ZrO_2$), [38] for tetragonal, and [39] for monoclinic $ZrO_2$. The surface cell of t-$ZrO_2$ deviates only slightly from two unit cells of c-$ZrO_2$.

## 2. Experimental

The UHV system used in this work comprises two-chambers, one for sample preparation and one for analysis. The preparation chamber (base pressure below $10^{-10}$ mbar) contains an ion source for sputtering, and an electron-beam heater for preparation of the substrate Rh(111) single crystal, as well as a home-built, UHV-compatible sputter source for deposition of Zr [36]. The thermocouple for temperature measurement is attached to the fixed part of the sample holder; temperatures above 700 °C are therefore corrected by using a disappearing-filament pyrometer and corrections for lower temperatures are extrapolated from the high temperature values. We estimate the temperatures to be accurate within ±30 °C. The analysis chamber ($p_{base} < 7 \times 10^{-11}$ mbar) houses a room-temperature STM (Omicron micro STM) and LEED optics (ErLEED). The whole system is suspended on springs for vibration damping. Etched W tips cleaned by $Ar^+$ sputtering were used in all STM measurements and conditioned by voltage pulses on a Au(110) crystal. All STM images showing atomic lattices or well-ordered superlattices were corrected for piezo drift as described in Ref. [29].

We chose Rh(111) as a substrate. Compared with Pt(111), it has the advantage of lower solubility of Zr in the bulk, and the 4:3 ratio of lattice constants between $ZrO_2$(111) and Rh favors the growth of unrotated zirconia films. A Rh(111) single crystal (diameter 9 mm, height 2 mm, from MaTecK, Germany) was cleaned by cycles of sputtering (2 keV $Ar^+$, 3.6 µA/cm², 10 min) and annealing (*T* = 920 °C,

10 min). Zirconium was sputter-deposited on the clean Rh substrate at RT in a mixed Ar/$O_2$ atmosphere ($p_{Ar}$ = 8 × $10^{-6}$ mbar, $p_{O_2}$ = 1 × $10^{-6}$ mbar); these films exhibit excellent purity [36]. The sputter deposition source also leads to some $Ar^+$ bombardment of the sample (ion-beam assisted deposition, IBAD); we chose rather gentle operating conditions with $Ar^+$ energies below 150 eV (grid voltages of 150 V and 100 V for the front and rear grid, respectively, unless noted otherwise) [36]. The amount of deposited material was calibrated by deposition of metallic Zr and measuring island areas with STM; the coverage was reproducible within 0.1 ML. We give the thickness in $ZrO_2$ monolayers (ML), with one O–Zr–O repeat unit of c-$ZrO_2$(111) defined as one monolayer, which corresponds to ≈ 9 × $10^{18}$ Zr atoms/$m^2$ or ≈ 0.3 nm thickness. The as-deposited films were not fully oxidized and were therefore post-annealed for 10 min in $O_2$ ($p_{O_2}$ = 5 × $10^{-7}$ mbar) at temperatures of at least 550 °C. In most experiments the post-annealing temperatures were such that a continuous but well-ordered film was obtained at the given film thickness; at higher temperatures and low film thickness (≤3 ML), holes down to the Rh substrate appeared.

## 3. Results
### 3.1. Zirconia layers of increasing thickness

When employing $ZrO_2$ films as a model system for the surface of bulk $ZrO_2$, a compromise between bulk-like properties (requiring thick films) and easy imaging by STM (requiring thin films due to their insulating nature) must be sought. Therefore, we have studied the structure of the films as a function of their thickness, starting from 1.5 ML. After annealing 1.5 ML of zirconia at 550 °C in $O_2$ ($p_{O_2}$ = 5 × $10^{-8}$ mbar), the film partially de-wetted the surface and a 2 ML-thick film with holes to the Rh(111) substrate formed, see Figure 2a. A single monolayer was found to be unstable under these annealing conditions. At first glance, the LEED pattern suggests a structure with (3 × 3) oxide units per (4 × 4) Rh cells (marked by red lines in Figure 2a), which would require an oxide lattice of 358 pm. This would be an expected structure, as three c-$ZrO_2$ unit cells (3 × 0.36 nm = 1.08 nm) nearly coincide with 4 Rh unit cells (4 × 0.269 nm = 1.076 nm). However, when using the Rh spots as a gauge to measure the true value of the oxide lattice constant, we find a value of approximately 0.34 nm – far shorter than the 358 pm required for a true (3 × 3)/(4 × 4) structure. The oxide structure can therefore not be explained by this superstructure. STM shows that the film is not perfectly ordered, as can be seen from the variations in the surface structure. In ordered areas, the most common feature resembles a rosette. The rosettes are hexagonally ordered with a periodicity of 1.2 nm, marked in the inset of Figure 2a. Usually, the domains of well-ordered rosettes are much smaller than in the inset of Figure 2a, however. From comparison with atomically resolved images of the Rh(111) surface (not shown), we find that the rosette lattice corresponds to a (√21 × √21)R10.9° superstructure with respect to Rh(111). We can explain the ideal rosette lattice using the lattice constant from LEED if it is a ($\sqrt{13} \times \sqrt{13}$)R13.9° superstructure w.r.t. a cubic $ZrO_2$(111) lattice. This results in an oxide lattice constant of 341 pm and a small 3° rotation of the oxide w.r.t. the Rh lattice.

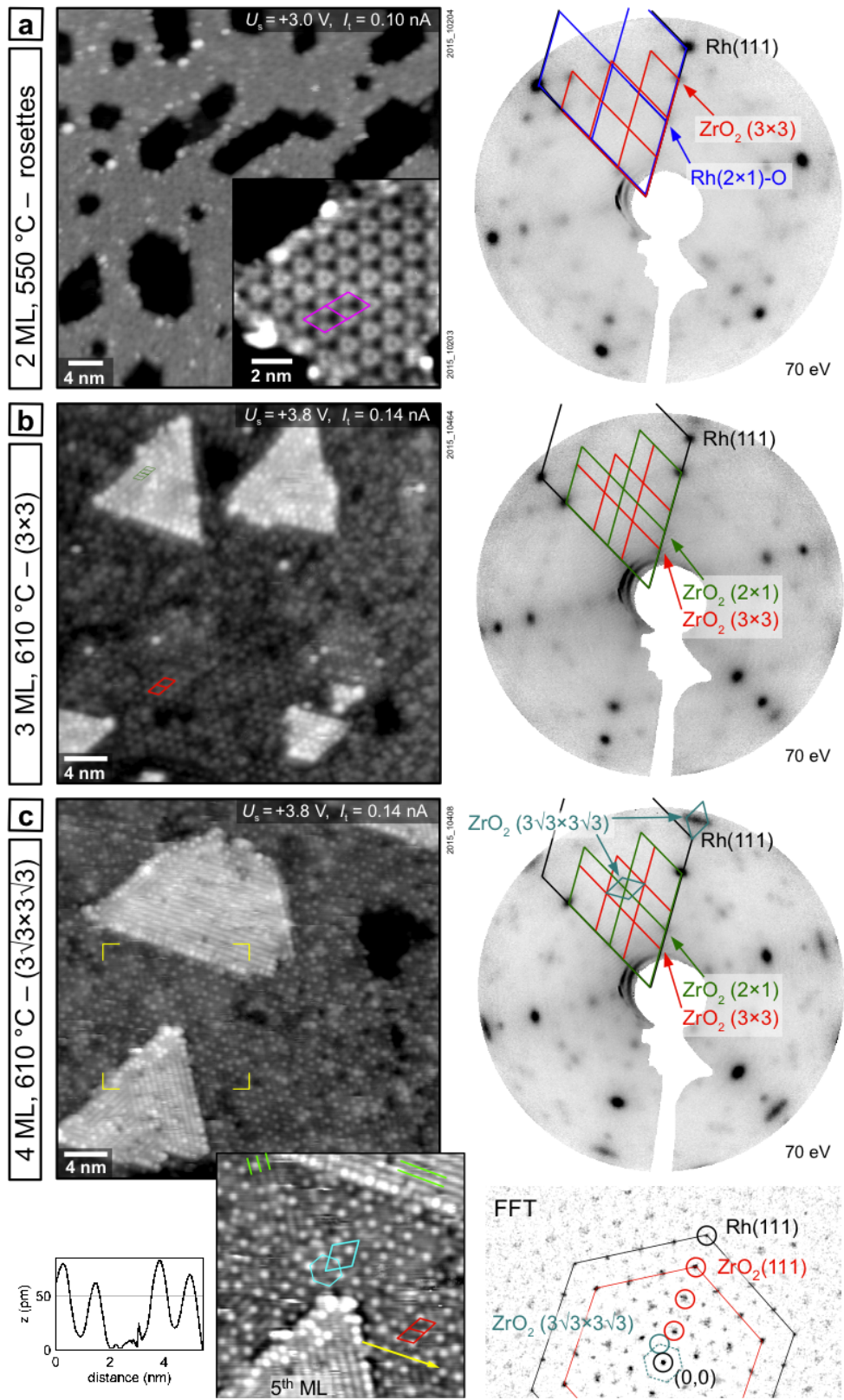

*Figure 2:* Surface structures of zirconia films with 2–4 ML thickness, as seen with STM (left) and LEED (right). Each thickness has its own surface reconstruction: Rosettes at 2 ML (a, unit cell marked in pink), (3 × 3) at 3 ML (b), and a (3√3 × 3√3) superstructure of small protrusions at 4 ML (c). For 4 ML, the line scan shows the height of the protrusions, and an FFT of the protrusions shows weak spots of their superstructure. Superstructures are given with respect to c-$ZrO_2$(111).

The in-plane lattice constant of 341 pm for $ZrO_2$ is surprisingly short: For metastable tetragonal or cubic $ZrO_2$ the corresponding value would be about 359–362 pm [38], and even 1 ML (single-trilayer) $ZrO_2$ films have a larger in-plane lattice constant of ≈350 pm [27,29] . As decreasing the lattice constant is constrained by O–O repulsion (see above), we consider it likely that these films are substantially oxygen-deficient. Based on the LEED image, however, the rotation of the oxide is less than 1° in most areas of the surface and therefore smaller than the 3° expected from the epitaxial relationship. This deviation can be explained, as we observe only small patches of well-ordered rosettes by STM, so the superstructure measured above is only an approximation. Thus, also the in-plane lattice constant may be slightly different from the one calculated assuming a perfectly commensurate superstructure.

We also observe LEED spots from a (2 × 1)-O structure on Rh(111) in the holes of the film. This structure is common when annealing Rh(111) in oxygen [40]; the corresponding periodicity can be also detected by STM in the holes (not shown).

A 3 ML-thick film annealed at 610 °C appears quite different in STM, see Figure 2b. Apart from some disordered regions in the upper half of the image, the predominant structure shows a (4 × 4) cell with respect to the substrate, which now nicely corresponds to (3 × 3) cells of the oxide (see the LEED pattern). This corresponds to an in-plane lattice constant of 358 pm, which is already close to the value for cubic zirconia (≈362 pm). There is no sign of the rosette structure that was found at 2 ML. It would be tempting to anneal to a higher temperature in order to improve the ordering and eliminate the disordered patches. Unfortunately, these thin films break up easily, forming thicker films with holes. These then have the structures of the respective thicker films. The 3 ML-thick film shown here already has a small number of holes down to the Rh substrate, which explains the bright Rh spots in LEED. In addition to the $ZrO_2$ (3 × 3) superstructure, LEED shows a $ZrO_2$ (2 × 1) structure that originates from 4 ML-high terraces (according to STM). This structure is discussed below.

In Figure 2c, a 4 ML-thick film is shown, with a few terraces having a total height of 5 ML. The LEED image again indicates a (3 × 3) oxide lattice per (4 × 4) Rh units. The 4 ML-thick areas are covered by isolated protrusions with a typical height of 60 pm, see the contrast-enhanced inset and the line profile in the inset. These protrusions can form a honeycomb-type short range order with a ($3\sqrt{3} \times 3\sqrt{3}$)R30° superstructure with respect to the oxide or ($4\sqrt{3} \times 4\sqrt{3}$)R30° with respect to Rh (hexagon in inset). We consider it likely that these protrusions are adatoms or molecules, but not impurities, as these features are solely present on the 4 ML films.

The protrusions nicely mark the lattice periodicity: A fast Fourier transform (FFT) of their positions extracted from the STM image (bottom right of Figure 2c) shows the Rh and $ZrO_2$ periodicities, as well as weaker spots for the $(3\sqrt{3} \times 3\sqrt{3})R30°$ lattice. Note that the circles in the FFT are exactly equidistant, marking the exact positions for a $(4 \times 4)$ superstructure. The oxide lattice is rotated by $\approx 0.5°$ with respect to the Rh lattice, which causes slight deviations of the maxima in the FFT from the center of the circles.

Between the protrusions, rows can be made out in the STM image in Figure 2c (marked by green lines in the top left of the inset). These rows have the same periodicity as the rows on the 5 ML-thick structure, see Figure 3a and the islands in Figure 2c. The distance between the rows is $\approx 0.6$ nm, which corresponds to a $(2 \times 1)$ structure with respect to the oxide, the expected unit cell of tetragonal zirconia. The 4 ML-high islands on the 3 ML film also show these rows (Figure 2b), thus the $(2 \times 1)$ already weakly appears in LEED at this coverage. This row structure persists also for thicker films, see below.

## 3.2. Tetragonal zirconia films

Zirconia films with a thickness of 5 ML and annealed at temperatures of up to 730 °C in $5 \times 10^{-7}$ mbar of $O_2$ are dominated by the row structure mentioned above. This structure has a $(2 \times 1)$ periodicity with respect to a $\approx 360$ pm c-$ZrO_2$(111) lattice, see Figure 3a. This is the structure expected for a tetragonal $ZrO_2$ film. As expected for the ABC stacking of t-$ZrO_2$, the rows of adjacent layers are laterally shifted by 1/3 of their spacing. This can be seen in panel a1 of Figure 3 (green lines). Domain sizes of $\approx 30$ nm can be reached upon annealing at 730 °C in $O_2$ (Figure 3a). The apparent corrugation of the tetragonal rows is surprisingly high (typically 30 pm; up to 100 pm at 2.4 V sample bias). This cannot be explained by the geometric heights of the surface atoms in the tetragonal structure ($\Delta z = 35$ pm for the O atoms in a bulk-terminated structure, less for a relaxed surface [16]; the Zr atoms have roughly equal heights). Thus, the high corrugation stems from either a surface reconstruction or an electronic effect. As neighboring domains of the tetragonal surface appear to blend into each other in some places, then appearing like a $(2 \times 2)$ structure (yellow circle in Fig. 3a), we consider the latter explanation more likely.

At a thickness of 5 ML, the surface structure can be atomically resolved with STM. With increasing film thickness, the bias voltage has to be increased for stable imaging, and the resolution of the images decreases accordingly (see the image of the 7.5 ML film in Figure 3b). It is difficult to obtain stable tunneling at a thickness of 10 ML; a minimum bias of 7.2 V is required. Nevertheless, the row structure of t-$ZrO_2$ remains visible at 7.5 ML (frame b1 of Figure 3) and at 10 ML (not shown), and the LEED pattern always shows a $(2 \times 1)$ pattern w.r.t. c-$ZrO_2$(111).

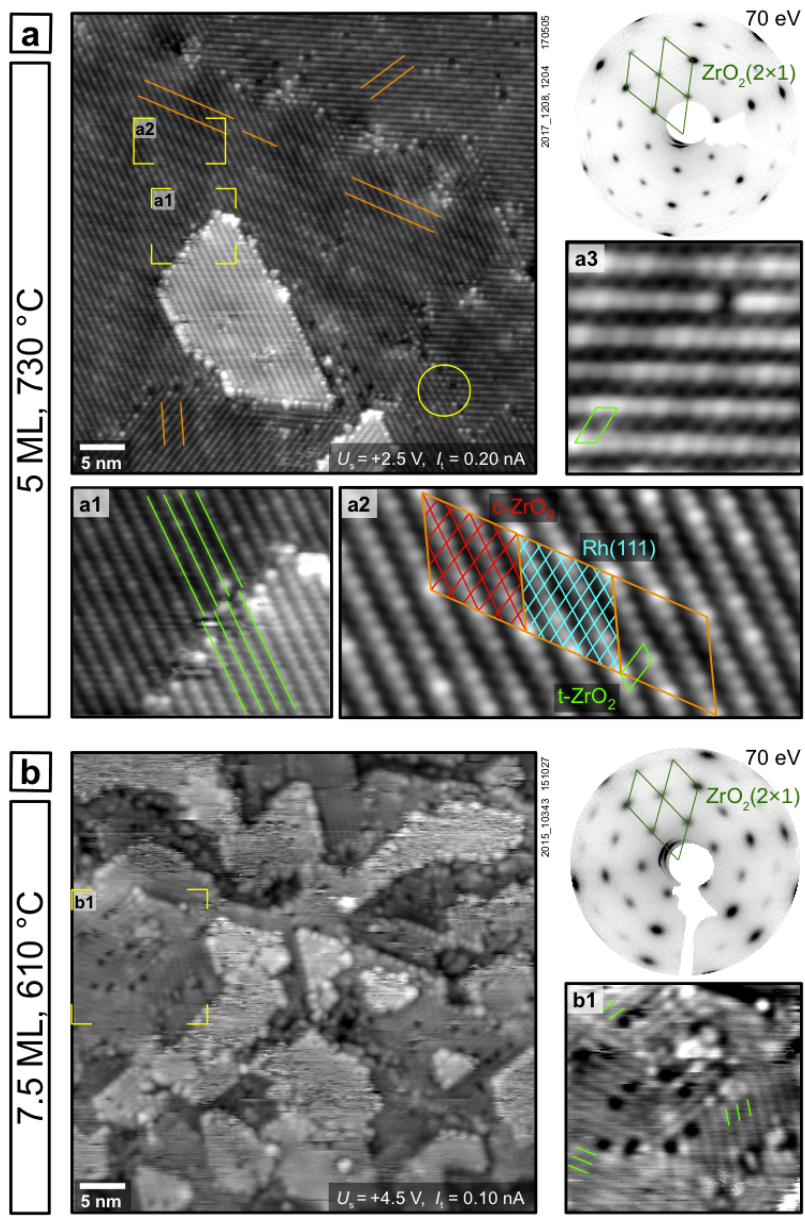

*Figure 3:* Tetragonal zirconia films with (a) 5 ML and (b) 7.5 ML, as seen with STM (large and zoom-in frames) and LEED (top right). Green lines mark the (2 × 1) surface structure with respect to cubic $ZrO_2$(111). When going from the 5th to the 6th ML, the rows shift by 1/3 of a unit cell as is expected for t-$ZrO_2$ (a1). Orange lines indicate the moiré structure visible in 5 ML films and some of its domain boundaries. The moiré superstructure cells are shown superimposed on the Fourier-filtered STM image in frame (a2), and a point defect is visible in frame (a3). The STM images have been processed to increase the contrast on the terraces.

Although an almost perfect 4:3 lattice match between tetragonal $ZrO_2$ and Rh(111) would be possible, the oxide is not exactly commensurate with the underlying Rh substrate. Upon careful inspection of the STM images, we find more than the three directions of the rows (in 120° increments) expected from the rotational symmetry of the substrate: The rows do not run exactly along the Rh $\langle\bar{1}10\rangle$ directions, but deviate from the close-packed directions of the substrate by up to ≈3°. This is accompanied by a moiré pattern, which is clearly visible in the STM images of the 5th ML (orange lines in Figure 3a). The moiré pattern becomes almost invisible in regions with 6 ML thickness and cannot be discerned in STM images of thicker films. There are several similar moiré patterns, however, and each type of moiré has six possible orientations of the oxide (three rotational domains, plus mirror symmetry). The different rotations of the zirconia film in different domains, which lead to the different moirés, cannot be resolved in LEED; rather than split into separate spots, the diffraction maxima of the hexagonal pattern in Figure 3 are only slightly elongated in the azimuthal direction.

For one of the domains with a nearly commensurate lattice, we could determine the moiré structure with respect to the substrate below (Figure 3, frame a2). This moiré cell corresponds to a $(7 \times 7)R21.8°$ superstructure with respect to Rh(111), which corresponds to $(2\sqrt{7} \times 2\sqrt{7})R19.1°$ cells of cubic $ZrO_2$(111), or half that number of tetragonal cells. This yields a rotation of 2.7° between the oxide lattice and the Rh(111) substrate; the average in-plane nearest-neighbor distance in the zirconia lattice is calculated as 355 pm and the in-plane angles between the nearest-neighbor directions would be multiples of exactly 60°. This moiré cell is only approximate, however. The moiré changes phase on a length scale of 10 nm; this can be seen at the orange lines in Figure 3a. The phase change probably happens because the interatomic distance of 355 pm would be too short for t-$ZrO_2$. In addition, this deviation from a perfectly commensurate cell also leads to a deviation from angles of exactly 60° (as expected for t-$ZrO_2$, see Figure 1b).

In other parts of the surface, we find roughly a 4:3 lattice match with the substrate in one direction, but nevertheless a moiré structure indicating a different (shorter) lattice constant in the other directions and deviations from 60° angles. Our best estimate for the average in-plane interatomic distances in the t-$ZrO_2$ films is around 357 pm, about 0.5% smaller than the room-temperature values from the literature [38], see Figure 1b.

## 3.3. Monoclinic zirconia films

Upon annealing a 5 ML-thick $ZrO_2$ film at 850 °C in $5 \times 10^{-7}$ mbar $O_2$, a phase transformation from t-$ZrO_2$ to m-$ZrO_2$ occurs. (Between 730 °C and 850 °C, the film is partially transformed.) Figure 4a shows a high-resolution STM image of the surface; the surface lattice appears hexagonal at first glance and no signs of the tetragonal row structure are visible. However, the monoclinic phase of zirconia is distorted with respect to the cubic and tetragonal phases, see Figure 1b. Due to this distortion, in order to compare the unit cell of our film with the cell size of bulk m-$ZrO_2$, we have to compare three different in-plane distances (or two distances plus one angle); approximate values for these three distances are shown in the inset of Figure 4a. In contrast to t-$ZrO_2$, the monoclinic lattice does not have an approximate 6-fold symmetry, which would help us correct for distortions of the STM images and thereby make an accurate determination of the lattice constants possible. As a way out, we took three sets of images with the fast scanning direction aligned with each of the $ZrO_2$ ⟨110⟩ directions. We then measured the distances along the fast scanning direction, which is almost unaffected by thermal drift or piezo creep. For calibration, we used atomically resolved images of the Rh(111) lattice recorded with the same scanning angle and scan speed (after removal of the oxide by sputtering). In our experience, this procedure should be accurate within ≈1–2%. The side lengths of the unit cell measured by this procedure are 727, 708, and 664 pm, which compares reasonably well with the values for m-ZrO$_2$($\bar{1}$11): 745, 733, and 678 pm [39]. The deviations from the expected values may hint at monoclinic distortions in the thin films being slightly different from the bulk. For comparison, the cell side lengths for the energetically less-favorable termination m-ZrO$_2$(111) [16] would be 797, 745, and 733 pm [39]. Thus, we can rule out the (111) orientation, which would be the only other symmetry-inequivalent type of m-ZrO$_2$\{111\} surfaces. The measured distances are also far from those expected for the orthorhombic structures; their unit cells have much less distortion with respect to c-ZrO$_2$. Thus, these films can be safely identified as m-ZrO$_2$($\bar{1}$11).

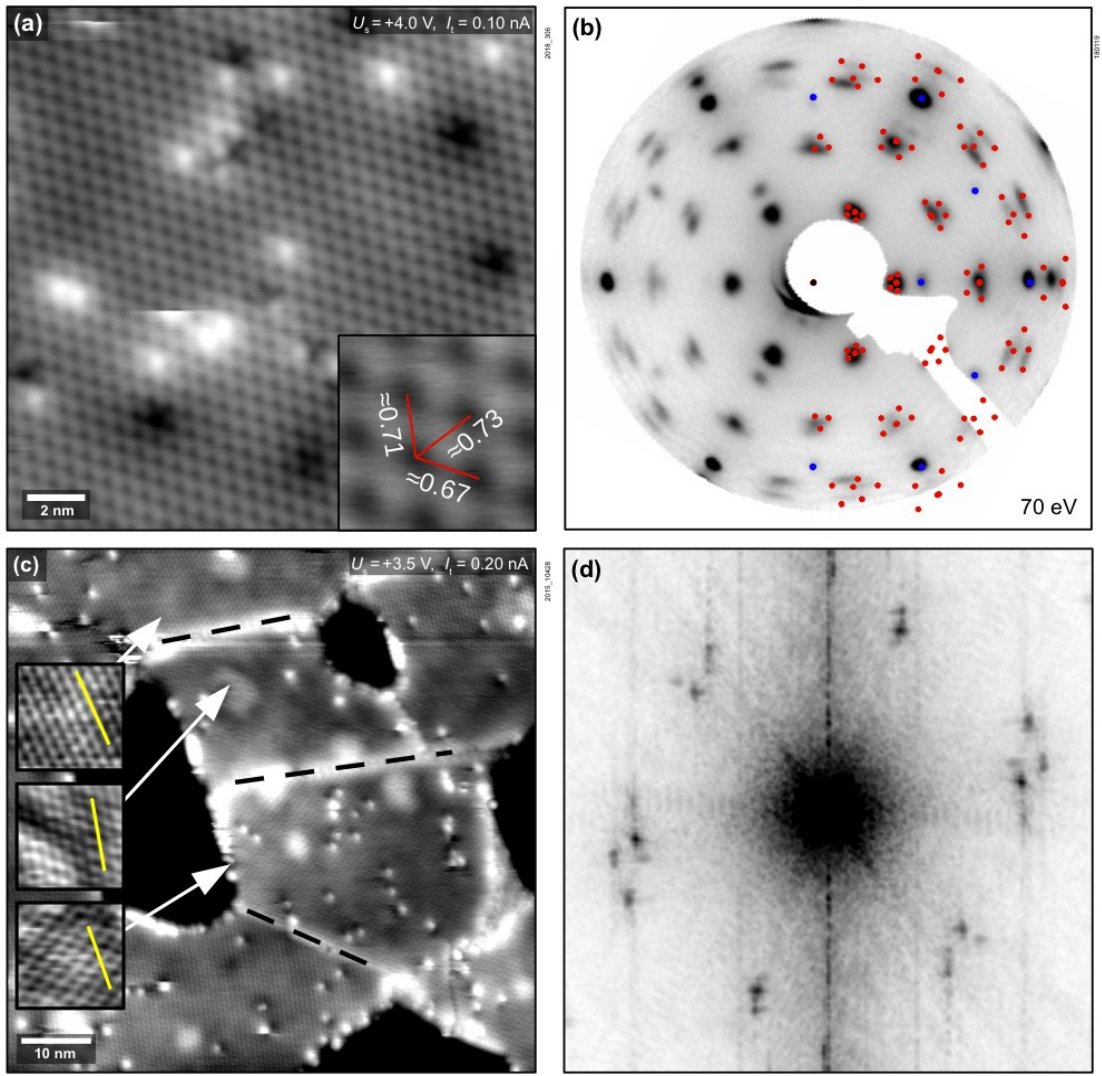

*Figure 4:* Monoclinic zirconia films: Upon annealing a 5 ML-thick film at 850 °C in $5 \times 10^{-7}$ mbar $O_2$, the film breaks up and transforms into the thermodynamically stable monoclinic structure. (a) High-resolution STM image of the structure. The inset shows a zoom to one unit cell, with approximate lattice parameters given in nm. (b) LEED image. The calculated LEED pattern of the $(\bar{1}11)$ surface of monoclinic zirconia is shown by red dots: spots originating from Rh(111) or Rh(111) (2 × 1)-O are blue. (c) STM overview image showing the holes down to Rh(111) and different domains (a few grain boundaries are marked by broken lines): In the Fourier transform, these domains result in a splitting of the spots (d).

In the FFT of an STM image with four different domains (Figure 4d), spots from the different domains can be seen in each direction. In LEED, these spots are smeared out, indicating slight variations in azimuthal orientation (Figure 4b). Nevertheless, the splitting of the monoclinic spots makes it easy to distinguish monoclinic and tetragonal films by LEED. Figure 4b also shows the expected diffraction pattern from six domains of m-ZrO$_2$($\bar{1}$11); these show a good agreement with experiment except for the right edge of the LEED screen, where the experimental image is distorted. Apart from the ZrO$_2$ spots, only Rh(111)-(1 × 1) and (2 × 1) spots are visible; the Rh(111)-(2 × 1) again stems from the (2 × 1)-O overlayer that is formed on bare Rh in the holes upon annealing in O$_2$ [40].

Together with the tetragonal-to-monoclinic transformation, the film usually breaks up, which leads to holes down to the Rh(111) substrate, see Figure 4c. The material from the holes spreads over the remaining zirconia areas, and locally increases the thickness (e.g., from 5 to 6 ML). The formation of holes and the phase transformation do not always go hand in hand, however: By changing the deposition parameters, we can prepare a tetragonal film that breaks up at temperatures below the phase transition point. In this case, we have lowered the front grid voltage of the sputter source from 150 V to 60 V, which reduced the energy of the Ar$^+$ ions that are hitting the surface during deposition [36]. Thus, the growth conditions become more comparable to thermal deposition. Such a weakly-sputtered film breaks up already at an annealing temperature of 670 °C, while it remains in the tetragonal structure. The phase transition then happens after annealing the film at 820 °C. The phase transformation from a tetragonal to a monoclinic film can be reversed by annealing at very high temperatures of 920 °C in UHV ($p_{base}$ < 1 × 10$^{-10}$ mbar). This preparation leads to a tetragonal film with holes down to Rh(111).

## 4. Discussion

Thin-film zirconia model systems have been studied since 1990 [25], yet the relation of these films to the ZrO$_2$ bulk structures could not be determined unambiguously. In hindsight, the LEED patterns originally interpreted as (2 × 2) with respect to c-

ZrO$_2$(111) by Maurice *et al.* and Meinel *et al.* [24,25] most likely correspond to three domains of the tetragonal structure, i.e., the three (2 × 1) domains in Figure 3. On Pt(111), these structures were not stable when annealing for more than one minute at 680 °C and transformed into (√19 × √19)R36.6° w.r.t. Pt(111) [24]; this is the same structure as the ultrathin zirconia films on Pt$_3$Zr(0001) [27]. It is possible that this low stability is due to easy dissolution of Zr in the Pt bulk (the dissolution enthalpy of Zr in Pt is exceptionally high [41,42]). On the other hand, the lower stability compared to our films might also be a consequence of thermal deposition in Ref. [22–24] vs. sputter deposition with additional ion bombardment in our case (remember that our films grown with softer ion bombardment than usual are less stable). The gentle Ar$^+$ ion bombardment by the sputter deposition source may help stabilizing the films by creating defects or slight intermixing; especially reactive Zr atoms embedded in Rh at the interface may help stabilizing the films by providing strong Zr–O bonds with O in the bottommost oxide layer (cf. the stabilization of zirconia films on Pd$_3$Zr(0001) [29]).

The transformation of the films to monoclinic zirconia was not reported in literature previously. It occurs at temperatures of 850 °C, so the higher thermal stability of our sputter-deposited films on Rh(111) compared to films created by thermal deposition on Pt(111) is certainly beneficial. Our attempts to obtain m-ZrO$_2$ by sputter deposition on Pt(111) were not successful when annealing in the same $p_{O2}$ = 5 × 10$^{-7}$ mbar as on Rh(111). The films broke up, but remained tetragonal up to 900 °C (not shown). However, at this temperature a significant part of the t-ZrO$_2$ film was reduced and Zr diffused into the Pt substrate. This does not happen on the Rh substrate. Starting from this partially reduced state on Pt(111), the transformation to m-ZrO$_2$ can be induced by annealing at 610 °C and higher $p_{O2}$ = 2 × 10$^{-6}$ mbar (not shown). Whether re-oxidation of dissolved Zr or the higher $p_{O2}$ is the reason for the stabilization of the monoclinic phase is a question for further studies.

Let us consider the stability of the bulk phases (Figure 1a), where m-ZrO$_2$ is stable at room temperature and t-ZrO$_2$ is the high-temperature phase. It is then surprising that the tetragonal phase in the thin films is stable at lower annealing temperatures ($T \leq 730$ °C) and transforms to the monoclinic phase when annealed at 850 °C. Assuming a lower surface energy for t-ZrO$_2$ than m-ZrO$_2$, it has been suggested that the monoclinic-to-tetragonal transition temperature decreases with decreasing film thickness (below 1 μm) and should reach room temperature in the range of 20 nm [43]. As mentioned in the introduction, the role of the surface energy stabilizing the tetragonal phase has to be questioned [12,13,15]. In any case, this is not the behavior encountered in our case, as the 5 ML (1.5 nm) tetragonal films can still be transformed to the monoclinic phase, which then stays stable upon cooling to RT. The tetragonal film is therefore in a metastable state, stabilized by the interface to the Rh substrate and, possibly, oxygen deficiency. Considering that Zr–O bonds get broken and the lattice gets distorted upon the tetragonal–to–monoclinic transformation (Fig. 1a), it is clear that a substantial activation energy is required to transform the films. The area density changes from 8.74 to 8.99 × $10^{18}$ Zr atoms m$^{-2}$ per layer (based on room-temperature bulk lattice constants [38,39]), and the transformation also involves in-plane shear. In contrast to the expansion of the interlayer spacing (295 to 317 pm), changing the area density and in-plane shear clearly require thermal activation.

When comparing the tetragonal and the monoclinic films on Rh(111), the tetragonal films appear rather flat, while the monoclinic films show long-distance modulations in their apparent height, with bright halos around both point defects and grain boundaries (Figure 4c). At positive STM sample bias, such an increase of the apparent height (increased tunneling current at constant height) is typical for downwards band bending [44]. These observations are important for the use of (chemically doped) zirconia as a solid-state electrolyte: Grain boundaries (GBs) in YSZ impede oxygen ion transport ("grain boundary blocking"); this is attributed to the positive charge at GBs [45], probably caused by oxygen vacancies at GBs (oxygen vacancies carry 2+ charge with respect to the undisturbed lattice with $O^{2-}$ at the respective site; $V_O^{\bullet\bullet}$ in Kröger-Vink notation [46]). Our STM images are consistent with this model.

The flat appearance of the tetragonal films points at a fixed position of the bands. Downwards band bending cannot occur if the conduction band minimum is close to the Fermi level, as is expected for a strongly n-doped oxide. This nicely fits the notion that oxygen vacancies (providing n doping) are responsible for the stabilization of the tetragonal phase in nanoscale $ZrO_2$, and could therefore be expected as a stabilizing factor in our thin films.

Let us finally discuss the films of lower thickness. While LEED indicates that all of these films are based on c-$ZrO_2$(111), the influence of the substrate does not allow these films to develop the surface structures expected for bulk $ZrO_2$. The in-plane lattice constants of the 2 ML film are clearly below those of the thicker films and the bulk phases. Again, this indicates sub-stoichiometric (oxygen-deficient) films; for example, it is possible that the rosette structure is related to ordering of oxygen vacancies. Starting from 3 ML, the in-plane lattice constants are already close to the bulk values (≈3:4 lattice match with Rh). For 4 ML films, the row structure of the tetragonal films is already locally visible by STM. It is unclear whether the bright features in Figure 2c are topographic (e.g. adatoms, small molecules, or clusters on top of the t-$ZrO_2$ surface) or electronic features. The high corrugation of these features (line scan in Figure 2c) points to a topographic feature. It is unlikely that these features are due to impurities, as they were not observed on thicker films.

## 5. Conclusion

Few-layer zirconia films can be reliably prepared by UHV-compatible sputter deposition. They show layer-by layer growth and good homogeneity. Up to a thickness of 4 ML, the surface structure of the zirconia films is influenced by the underlying Rh(111) substrate. For each thickness below 5 ML, a different superstructure is found; apart from the 2 ML films, all structures are close to a commensurate lattice with $(3 \times 3)$ c-$ZrO_2$(111) cells on $(4 \times 4)$ Rh(111) unit cells. Zirconia films of 5 ML or larger thickness show the surface structures of either tetragonal or monoclinic zirconia, depending on the annealing temperature. Both structures can be prepared with large, atomically flat terraces; their surface lattices were resolved by STM, confirming their crystallographic structure. Preparation of a completely monoclinic film needs annealing at temperatures of at least 850 °C; at these temperatures, the film breaks up and holes reaching down to the substrate appear. Thus, the films show some instability towards dewetting. Due to the insulating nature of $ZrO_2$, imaging the surface with STM becomes increasingly difficult with increasing film thickness, thus the thinnest films showing the structures of the $ZrO_2$ bulk phases (5 ML) are the best choice for an STM-accessible $ZrO_2$ model system. In comparison with the previous $ZrO_2$/Pt(111) model system, we believe that there are three reasons for the superior film homogeneity and stability on Rh(111): Firstly, the lower solubility of Zr in Rh than in Pt. Secondly, the 3:4 lattice matching, which leads to low rotation angles between different domains, and thirdly the use of a UHV-compatible sputter source providing additional slight ion bombardment.

## Acknowledgements

The authors want to thank Bilge Yildiz (MIT, USA) and Jürgen Fleig (TU Wien, Austria) for helpful discussions, and Andreas Stierle and Vedran Vonk (DESY, Germany) for x-ray diffraction measurements providing additional confirmation of the tetragonal structure of a 10 ML t-$ZrO_2$ film. This work has been supported by the Austrian Science Fund (FWF) under project numbers F4505 and F4507.

## References

[1] T. Yamaguchi, Catal. Today **20**, 199 (1994).
[2] K. Tanabe, Mater. Chem. Phys. **13**, 347 (1985).
[3] R. H. J. Hannink, P. M. Kelly, and B. C. Muddle, J. Am. Ceram. Soc. **83**, 461 (2000).
[4] R. C. Garvie, R. H. Hannink, and R. T. Pascoe, Nature **258**, 703 (1975).
[5] N. P. Padture, M. Gell, and E. H. Jordan, Science **296**, 280 (2002).
[6] P. F. Manicone, P. Rossi Iommetti, and L. Raffaelli, J. Dent. **35**, 819 (2007).
[7] S. McIntosh and R. J. Gorte, Chem Rev **104**, 4845 (2004).
[8] Y. Liu, J. Parisi, X. Sun, and Y. Lei, J. Mater. Chem. A **2**, 9919 (2014).
[9] E. H. Kisi and C. J. Howard, Key Eng. Mater. **153**, 1 (1998).
[10] S. P. Terblanche, J. Appl. Crystallogr. **22**, 283 (1989).
[11] M. Hillert and T. Sakuma, Acta Metall. Mater. **39**, 1111 (1991).
[12] S. Shukla and S. Seal, Int. Mater. Rev. **50**, (2005).

[13] M. Raza, D. Cornil, J. Cornil, S. Lucas, R. Snyders, and S. Konstantinidis, Scr. Mater. **124**, 26 (2016).
[14] H. G. Scott, J. Mater. Sci. **10**, 1527 (1975).
[15] R. C. Garvie, J. Phys. Chem. **69**, 1238 (1965).
[16] A. Christensen and E. A. Carter, Phys. Rev. B **58**, 8050 (1998).
[17] J. M. Leger, P. E. Tomaszewski, A. Atouf, and A. S. Pereira, Phys. Rev. B **47**, 14075 (1993).
[18] J. Müller, T. S. Böscke, U. Schröder, S. Mueller, D. Bräuhaus, U. Böttger, L. Frey, and T. Mikolajick, Nano Lett. **12**, 4318 (2012).
[19] R. Materlik, C. Künneth, and A. Kersch, J. Appl. Phys. **117**, 134109 (2015).
[20] O. Ohtaka, T. Yamanaka, S. Kume, N. Hara, H. Asano, and F. Izumi, Proc. Jpn. Acad. **66**, 193 (1990).
[21] S. L. Morrow, T. Luttrell, A. Carter, and M. Batzill, Surf. Sci. **603**, L78 (2009).
[22] K. Meinel, K.-M. Schindler, and H. Neddermeyer, Surf. Sci. **532–535**, 420 (2003).
[23] K. Meinel, A. Eichler, K.-M. Schindler, and H. Neddermeyer, Surf. Sci. **562**, 204 (2004).
[24] K. Meinel, A. Eichler, S. Förster, K.-M. Schindler, H. Neddermeyer, and W. Widdra, Phys. Rev. B **74**, 235444 (2006).
[25] V. Maurice, M. Salmeron, and G. A. Somorjai, Surf. Sci. **237**, 116 (1990).
[26] Y. Gao, L. Zhang, Y. Pan, G. Wang, Y. Xu, W. Zhang, and J. Zhu, Chin. Sci. Bull. **56**, 502 (2011).
[27] M. Antlanger, W. Mayr-Schmölzer, J. Pavelec, F. Mittendorfer, J. Redinger, P. Varga, U. Diebold, and M. Schmid, Phys. Rev. B **86**, 035451 (2012).
[28] H. Li, J.-I. J. Choi, W. Mayr-Schmölzer, C. Weilach, C. Rameshan, F. Mittendorfer, J. Redinger, M. Schmid, and G. Rupprechter, J. Phys. Chem. C **119**, 2462 (2015).
[29] J. I. J. Choi, W. Mayr-Schmölzer, F. Mittendorfer, J. Redinger, U. Diebold, and M. Schmid, J. Phys. Condens. Matter **26**, 225003 (2014).
[30] J. I. J. Choi, W. Mayr-Schmölzer, I. Valenti, P. Luches, F. Mittendorfer, J. Redinger, U. Diebold, and M. Schmid, J. Phys. Chem. C **120**, 9920 (2016).
[31] P. Lackner, J. Hulva, E.-M. Köck, W. Mayr-Schmölzer, J. I. J. Choi, S. Penner, U. Diebold, F. Mittendorfer, J. Redinger, B. Klötzer, G. S. Parkinson, and M. Schmid, Submitted (2018).
[32] J. P. Chang, Y.-S. Lin, S. Berger, A. Kepten, R. Bloom, and S. Levy, J. Vac. Sci. Technol. B Microelectron. Nanometer Struct. Process. Meas. Phenom. **19**, 2137 (2001).
[33] L. Mayr, X.-R. Shi, N. Köpfle, C. A. Milligan, D. Y. Zemlyanov, A. Knop-Gericke, M. Hävecker, B. Klötzer, and S. Penner, Phys. Chem. Chem. Phys. **18**, 31586 (2016).
[34] N. Köpfle, L. Mayr, P. Lackner, M. Schmid, D. Schmidmair, T. Götsch, S. Penner, and B. Klötzer, ECS Trans. **78**, 2419 (2017).
[35] L. Mayr, X. Shi, N. Köpfle, B. Klötzer, D. Y. Zemlyanov, and S. Penner, J. Catal. **339**, 111 (2016).
[36] P. Lackner, J. I. J. Choi, U. Diebold, and M. Schmid, Rev. Sci. Instrum. **88**, 103904 (2017).
[37] D. W. Strickler and W. G. Carlson, J. Am. Ceram. Soc. **48**, 286 (1965).
[38] N. Igawa and Y. Ishii, J. Am. Ceram. Soc. **84**, 1169 (2001).
[39] C. J. Howard, R. J. Hill, and B. E. Reichert, Acta Crystallogr. B **44**, 116 (1988).
[40] M. V. Ganduglia-Pirovano and M. Scheffler, Phys. Rev. B **59**, 15533 (1999).
[41] Y. Gao, C. Guo, C. Li, and Z. Du, Int. J. Mater. Res. Former. Z. Fuer Met. **101**, 819 (2010).
[42] A. K. Niessen and A. R. Miedema, Berichte Bunsenges. Für Phys. Chem. **87**, 717 (1983).
[43] J. Müller, T. S. Böscke, U. Schröder, M. Reinicke, L. Oberbeck, D. Zhou, W. Weinreich, P. Kücher, M. Lemberger, and L. Frey, Microelectron. Eng. **86**, 1818 (2009).
[44] Ebert Ph., Surf. Sci. Rep. **33**, 121 (1999).
[45] X. Guo, W. Sigle, J. Fleig, and J. Maier, Solid State Ion. **154–155**, 555 (2002).
[46] O. Yamamoto, Electrochim. Acta **45**, 2423 (2000).